\documentclass[a4paper,11pt]{article}
\usepackage{pos}

\title{Nucleon transverse quark spin densities}

\author*[a,b]{Constantia Alexandrou}
\author[b]{Simone Bacchio}
\author[c]{Martha Constantinou}
\author[d]{Petros Dimopoulos}
\author[b]{Jacob Finkenrath}
\author[e]{Roberto Frezzotti}
\author[a,b]{Kyriakos Hadjiyiannakou}
\author[f]{Karl Jansen}
\author[g]{Bartosz Kostrzewa}
\author[b]{Giannis Koutsou}
\author[b]{Gregoris Spanoudes}
\author[h,i]{Carsten Urbach}

\affiliation[a]{Department of Physics, University of Cyprus, P.O. Box 20537, 1678 Nicosia, Cyprus}

\affiliation[b]{Computation-based Science and Technology Research Center, The Cyprus Institute, 20 Kavafi Street, Nicosia 2121, Cyprus}
\affiliation[c]{Department of Physics, Temple University, Philadelphia, PA 19122 - 1801, USA}
\affiliation[d]{Dipartimento di Scienze Matematiche, Fisiche e Informatiche, Universit\`a di Parma and INFN,\\ Gruppo Collegato di Parma Parco Area delle Scienze 7/a (Campus), 43124 Parma, Italy}
\affiliation[e]{Dipartimento di Fisica and INFN, Universit\'a di Roma Tor Vergata, Via della Ricerca Scientifica 1, I-00133 Roma, Italy}
\affiliation[f]{NIC, DESY Zeuthen, Germany}
\affiliation[g]{High Performance Computing and Analytics Lab, Rheinische Friedrich-Wilhelms-Universit\"at Bonn,\\
Friedrich-Hirzebruch-Allee 8, 53115 Bonn, Germany}
\affiliation[h]{Bethe Center for Theoretical Physics, University of Bonn, Germany}
\affiliation[i]{Helmholtz-Institut f\"ur Strahlen- und Kernphysik, University of Bonn, Germany}

\emailAdd{alexand@ucy.ac.cy}

\abstract{We present a calculation of the  Mellin moments of the nucleon transverse quark spin densities extracted from the unpolarized and transversity generalized form factors. We use three $N_F=2+1+1$ ensembles of twisted mass fermions with quark masses tuned to their physical values and lattice spacings $a\sim 0.08$~fm,  $a\sim 0.07$~fm and $a\sim 0.06$~fm and extrapolate the form factors to the continuum limit. Besides isovector densities we also include results for the tensor charge for each quark flavor using the ensemble with $a\sim 0.08$~fm for which   we include the disconnected contributions.  }

\FullConference{%
 The 39th International Symposium on Lattice Field Theory, LATTICE2022
  8th-13th August, 2022,
 University of  Bonn, Germany
}

\begin{document}
\maketitle

\section{Introduction}
 Understanding  the 3D-structure of the nucleon from its fundamental constituents, the quarks and the gluons is a  major goal of nuclear physics and a key aim of current and future major experiments. 
Significant progress has been made  in revealing the longitudinal spin  structure of the proton. 
However, the  transverse spin structure is much less known and will be the focus of the experimental programs of  Jefferson Lab and the future Electron Ion Collider(EIC)  being built at Brookhaven~\cite{Accardi:2012qut}. Although recently techniques to compute the  generalized parton distributions (GPDs)  using lattice QCD are being developed, computation of moments of GPDs is at a more mature stage.
In particular, lower moments  are readily accessible from  lattice QCD by computing  matrix elements of local operators.
 Although the computation of the low Mellin moments has a long history~\cite{Martinelli:1988xs}, it is only  recently that results directly at the physical point are emerging (i.e. using ensembles generated with  pion mass at the physical point defined as $m_\pi = 135\pm10$~ MeV).
 To determine the transverse spin densities, one needs to compute the twist-two  chiral-even unpolarized and chiral-odd transversity GPDs.
 \section{Nucleon matrix elements}

 The transverse spin density is  defined as~\cite{Diehl:2005jf}
 \begin{eqnarray}
  && \rho(x,{\bf b}_\perp,{\bf s}_\perp,{\bf S}_\perp) = \frac{1}{2} \bigg[ H(x,b^2_\perp) + \frac{{\bf b}_\perp^j \epsilon^{ji}}{m_N} \left( {\bf S}_\perp^i E'(x,b^2_\perp) + {\bf s}_\perp^i \bar{E}'_{T}(x,b^2_\perp)\right) \nonumber \\
     &+& {\bf s}_\perp^i {\bf S}_\perp^i \left( H_T(x,b^2_\perp) - \frac{\Delta_{b_\perp} \widetilde{H}_T(x,b^2_\perp)}{4 m_N^2} \right) +{\bf s}_\perp^i(2{\bf b}_\perp^i {\bf b}_\perp^j- \delta^{ij} b^2_\perp){\bf S}_\perp^j \frac{\widetilde{H}''_T(x,b^2_\perp)}{m_N^2},
   \label{rho}
 \end{eqnarray}
 where the generalized parton distributions $H, E, H_T , E_T$ and $\widetilde{H}_T$ in Eq.~(\ref{rho}) are
given in impact parameter space for zero skewness. They are obtained  
by a Fourier transformation, ${\bf \Delta}_\perp \rightarrow {\bf  b}_\perp$, ${\bf \Delta}_\perp$
is the transverse momentum transfer and $-Q^2 \equiv \Delta^2$. 
$m_N$ is the nucleon mass, $x$ is the longitudinal momentum fraction, ${\bf s}_\perp$ is  the transverse quark spin, ${\bf b}_\perp$  the transverse vector from the center of momentum of the nucleon, and ${\bf S}_\perp$ the transverse spin of the nucleon, $\epsilon_{ij}$  is the antisymmetric tensor, the derivatives are denoted as $F^\prime\equiv \frac{\partial F}{\partial b^2_\perp}$ and $\Delta_{b_\perp} F\equiv=4\frac{\partial}{\partial b^2}\left({\bf b}_\perp^2\frac{\partial}{\partial b^2}\right)F$ and the GPD $\bar{E}_T=E_T+2\widetilde H_T$.
The moments of $\rho$ given by
\begin{equation}
  \langle x^{n-1}\rangle_\rho({\bf b}_\perp,{\bf s}_\perp, {\bf S}_\perp)\equiv \int_{-1}^1 \, dx\, x^{n-1} \rho(x,{\bf b}_\perp,{\bf s}_\perp, {\bf S}_\perp),
  \label{rho moments}
\end{equation}
are related to generalized form factors (GFFs), e.g.
 $ A_{Tn0}=\int_{-1}^1\,dx\,x^{n-1}H_T$,  $B_{Tn0}=\int_{-1}^1\, dx \, x^{n-1}E_T$, $\widetilde{A}_{Tn0}=\int_{-1}^1\,dx\,x^{n-1} \widetilde{H}_{T}$.
The relevant GFFs are determined from the two lowest Mellin moments by considering the  nucleon matrix elements of the following operators:\\
$\bullet$ Tensor operator, yielding the first Mellin  tensor GFFs
 \begin{eqnarray}
 & &\langle N (p^\prime,s^\prime)| \bar{q} \sigma^{\mu\nu} q| N (p,s)\rangle =
\bar{u}_N(p^\prime,s^\prime)\left[\sigma^{\mu\nu}\, A^q_{T10}(Q^2)
+ i \frac{\gamma^{[\mu}\Delta^{\nu]}} {2 m_N} \right. \,B^q_{T10}(Q^2)\nonumber \\
&+& \left. \frac{\overline P^{[\mu} \Delta^{\nu]}} {m_N^2 }\, \widetilde A^q_{T10}(Q^2) \right]u_N(p,s),
\label{tensor}
  \end{eqnarray}
  where with $[...]$ we denote antisymmetrization  over the indices and $ \bar{P}^\mu=\frac{p^{\prime\mu}+p^\mu}{2}$.
  In the forward limit,  $A_{T10}(0)$ gives the tensor charge $g_T$, while $\bar{B}^q_{T10}(0) \equiv B^q_{T10}(0)+2\widetilde{A}^q_{T10}(0)$ is the anomalous tensor magnetic moment $\kappa_T$, which is related to the Boer-Mulders function $h^\perp\sim -\kappa_T$.\\
$\bullet$ Vector one-derivative operator $\mathcal{O}^{\mu\nu}_{\rm V} = \bar{q} \gamma^{\{\mu}iD^{\nu\}} q$, yielding the unpolarized second Mellin GFFs
  \begin{eqnarray}
 & &  \langle N(p',s') \vert \mathcal{O}^{\mu\nu}_{\rm V} \vert N(p,s) \rangle = \bar{u}_N(p',s') \Big[ A_{20}(Q^2) \gamma^{\{\mu }\overline P^{\nu\}} + B_{20}(Q^2) \frac{i\sigma^{\{\mu\alpha} q_\alpha \overline P^{\nu\}}}{2m_N} \nonumber \\
 &+& C_{20}(Q^2) \frac{q^{\{\mu} q^{\nu\}}}{m_N} \Big] u_N(p,s),
    \label{unpolarized}
  \end{eqnarray}
  where  with $\{...\}$ we denote symmetrization over the indices and subtraction of the trace.
  In the forward limit, $A_{20}(0)$  gives the quark momentum fraction $\langle x\rangle_q$ and the total quark spin is $J_q=\frac{1}{2}\left[A_{20}(0)+B_{20}(0)\right]$.\\
  $\bullet$  Tensor one-derivative operator ${\cal O}_T^{\mu\nu\rho} = i\, \bar{q}\sigma^{[\mu\{\nu]}\,i\,D^{\rho\}}\,q$, yielding the second Mellin tensor  GFFs with the transversity moment ${\langle  x \rangle_{\delta q} = A^q_{T20}(0)}$.
    \begin{eqnarray}
     & & \hspace*{-1.7cm}\langle N(p^\prime,s^\prime)\vert{\cal{O}}_T^{\mu\nu\rho} \vert N(p,s)\rangle =
u_N(p^\prime,s^\prime)\left[i \sigma^{\mu\nu} \overline P^{\rho}\, A_{T20}(Q^2)
+ \frac{\gamma^{[\mu}\Delta^{\nu]}} {2 m_N} \overline P^{\rho} \,B_{T20}(Q^2)\,\right. \nonumber \\
&+&   i\,\frac{\overline P^{[\mu} \Delta^{\nu]}} {m_N^2 } \overline P^{\rho}\, \widetilde A_{T20}(Q^2)+\left.\frac{\gamma^{[\mu}\overline P^{\nu]}} {m_N} \Delta^{\rho} \, \widetilde B_{T21}(Q^2) \right]u_N(p,s).
\label{tensor 2nd}
    \end{eqnarray}
\section{Results}

\begin{table}
  \caption{The first column gives the ensemble name, the second the lattice volume, the third the lattice spacing, the fourth the value of the pion mass $m_\pi$, the fifth the spatial length in fm and the last  $m_\pi L$.\label{tab:params}}
\begin{tabular}{||c||c|c|c|c||c|c||}
    \hline
    ~~~ ensemble name~~~  & ~~~ $(L/a)^3.T/a$ ~~~ & ~~~ $a$ (fm) ~~~ & ~~~ $m_{\pi}$ (MeV) ~ & ~ $L$ (fm) ~ & $m_{\pi}L$ ~ \\
  \hline
  cB211.072.64 &  $64^{3}\cdot 128$ & $0.07961~(13)$ &  $140.2~(2)$ & $5.09$ & $3.62$ \\
  
  
  cC211.060.80 & $80^{3}\cdot 160$ & $0.06821~(12)$  & $136.7~(2)$ & $5.46$ & $3.78$ \\
  
  cD211.054.96 &  $96^{3}\cdot 192$ & $0.05692~(10)$  & $140.8~(2)$ & $5.46$ & $3.90$ \\
  \hline
\end{tabular}
\end{table}

\begin{figure}[h!]
  \includegraphics[width=\linewidth]{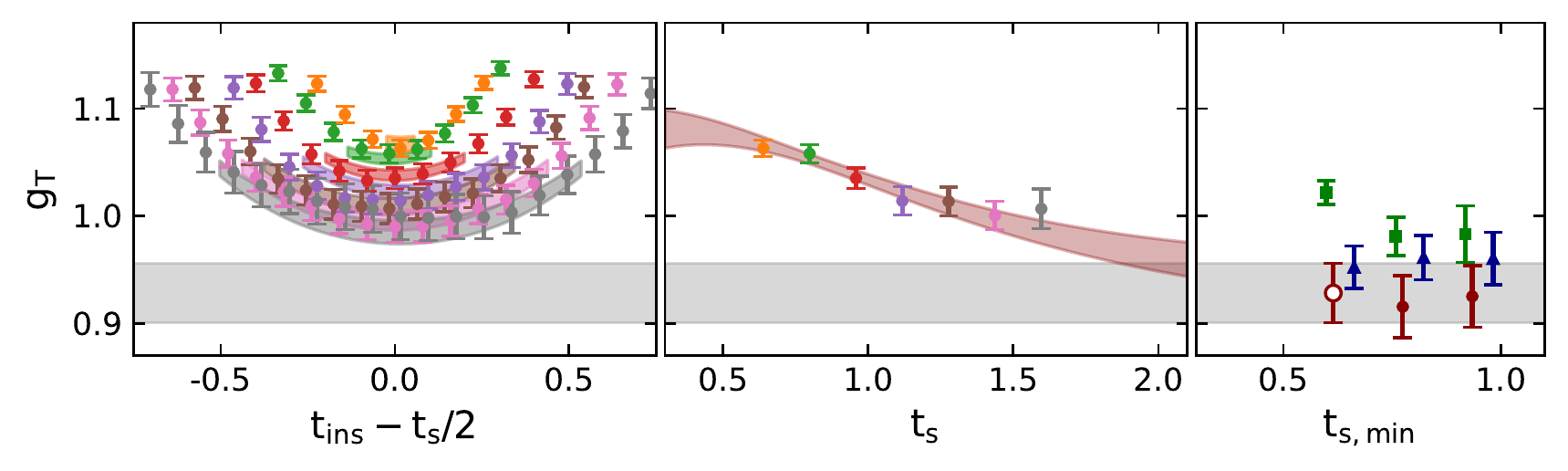}\\
  \includegraphics[width=\linewidth]{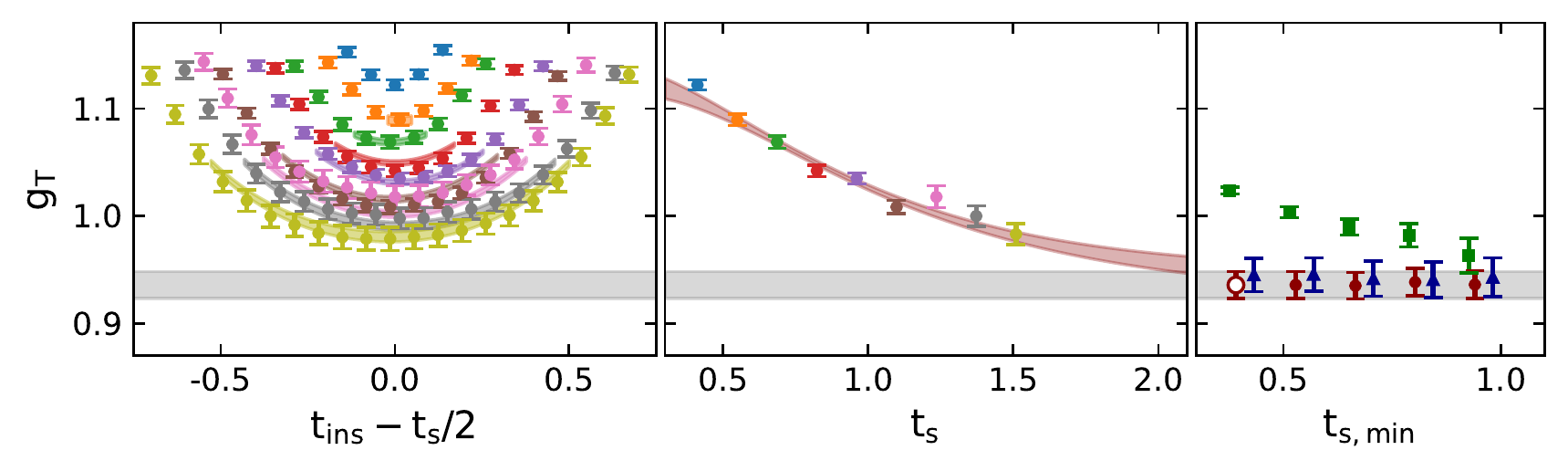}\\
  \includegraphics[width=\linewidth]{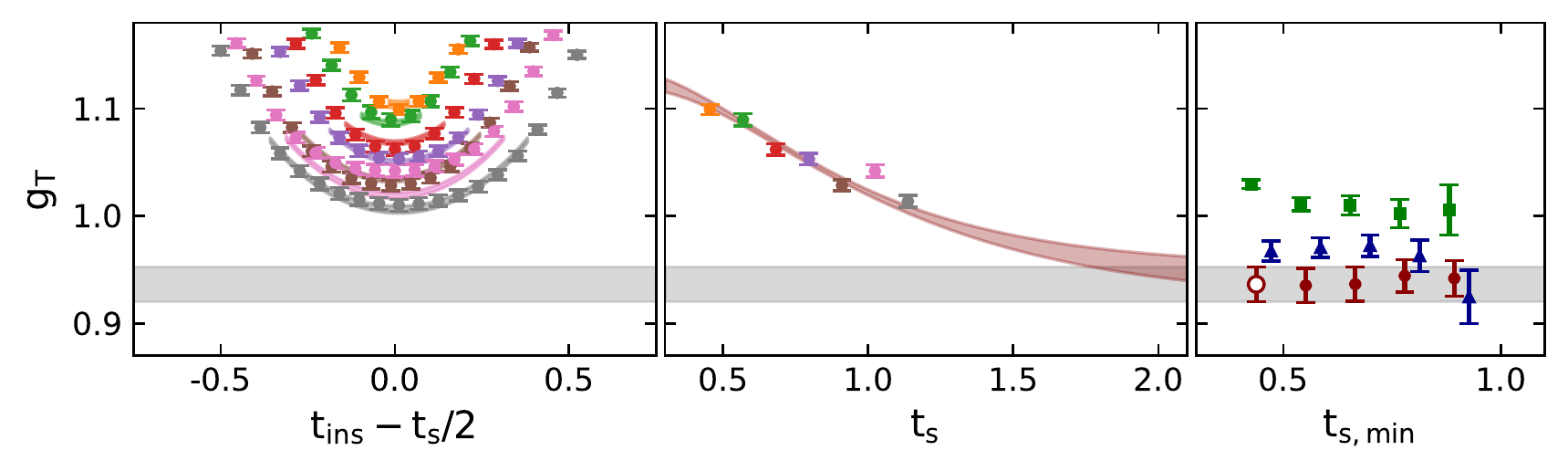}
  \caption{We show the analysis of the ratio  from which we extract $g_T^{u-d}$. The left panels show the ratio versus the insertion time $t_{\rm ins}$ shifted by half the source-sink time $t_s/2$. The middle panels show the values extracted when fitting the ratio on the left with a constant leaving $8a$ time separation from the source and sink, or the value of the central point when not enough values of $t_s$  are available. In the right panel, we show the values obtained when we perform a two state-fit (red circles),  a three-state fit (blue triangles) and the summation method (green circles), as we vary the lower fit range, $t_{s,{\rm min}}$  used in the fits. The open symbol shows the value we take and the red band in the middle panels shows the prediction of the plateau values for a given  $t_s$ value using the selected parameters determined by the open symbols. Results are shown for  the cB211.072.64 (top), cC211.060.80 (middle) and cD211.054.96 (bottom) ensemble. For the latter, we are still increasing statistics for the three largest values of $t_s$, see Table.~\ref{tab:statistics}.}
  \label{fig:gT excited}\vspace*{-0.5cm}
  \end{figure}
 We illustrate our analysis by considering the extraction of $g_T$ for the three $N_f=2+1+1$ twisted mass clover-improved fermion ensembles simulated by the Extended Twisted Mass Collaboration (ETMC) using physical values of the two light, the strange and the charm quark masses. The parameters of the ensembles are listed in Table~\ref{tab:params}. 
  In Fig.~\ref{fig:gT excited}, we show the extraction of the isovector $g_T^{u-d}$ using three physical listed in Table~\ref{tab:params}. As can be seen, there are large excited state contributions that would require $t_s>2$~fm to eliminate in the ratio~\cite{Bar:2016uoj}. However, including a the first excited state, the results  converge rapidly yielding the nucleon matrix element. This is confirmed by including the second excited state in the fit. Note that having approximately the same statistical error as we increase $t_s$ is important for identifying clearly the nucleon matrix element. The summation method, on the other hand, shows  slow  convergence, but eventually converges to the chosen value  from the two-state fit. In Table~\ref{tab:statistics} we give the statistics used.
  \begin{table}[h!]
    \begin{minipage}[t]{0.33\linewidth}
    \centering
    \begin{tabular}{|r|r|r|}
    \hline
      \multicolumn{3}{|c|}{\bf cB211.072.64} \\
    \hline
      \multicolumn{3}{|c|}{750 configurations} \\
    \hline
      $t_s/a$ & $t_s$[fm] & $n_{src}$ \\
    \hline
      8 & 0.64 &  1 \\
     10 & 0.80 &  2 \\
     12 & 0.96 &  4 \\
     14 & 1.12 &  6 \\
     16 & 1.28 & 16 \\
     18 & 1.44 & 48 \\
     20 & 1.60 & 64 \\
    \hline
     \multicolumn{2}{|c|}{Nucleon 2pt} & 264 \\
    \hline
    \multicolumn{3}{}{} \\
    \multicolumn{3}{}{} \\
    \multicolumn{3}{}{} \\
    \end{tabular}
   \end{minipage}
   \begin{minipage}[t]{0.33\linewidth}
    \centering
    \begin{tabular}{|r|r|r|}
    \hline
      \multicolumn{3}{|c|}{\bf cC211.060.80} \\
    \hline
      \multicolumn{3}{|c|}{400 configurations} \\
    \hline
      $t_s/a$ & $t_s$[fm] & $n_{src}$ \\
    \hline
      6 & 0.41 &   1 \\
      8 & 0.55 &   2 \\
     10 & 0.69 &   4 \\
     12 & 0.82 &  10 \\
     14 & 0.96 &  22 \\
     16 & 1.10 &  48 \\
     18 & 1.24 &  45 \\
     20 & 1.37 & 116 \\
     22 & 1.51 & 246 \\
    \hline
     \multicolumn{2}{|c|}{Nucleon 2pt} & 650 \\
    \hline 
    \multicolumn{3}{}{} \\
    \end{tabular}
  \end{minipage}
  \begin{minipage}[t]{0.33\linewidth}
    \centering
    \begin{tabular}{|r|r|r|}
    \hline
      \multicolumn{3}{|c|}{\bf cD211.054.96} \\
    \hline
      \multicolumn{3}{|c|}{500 configurations} \\
    \hline
      $t_s/a$ & $t_s$[fm] & $n_{src}$ \\
    \hline
      8 & 0.46 &   1 \\
     10 & 0.57 &   2 \\
     12 & 0.68 &   4 \\
     14 & 0.80 &   8 \\
     16 & 0.91 &  16 \\
     18 & 1.03 &  32 \\
     20 & 1.14 &  64 \\
     22 & 1.25 &  16 \\
     24 & 1.37 &  32 \\
     26 & 1.48 &  64 \\
    \hline
     \multicolumn{2}{|c|}{Nucleon 2pt} & 368 \\
    \hline 
    \end{tabular}
  \end{minipage}
      \caption{Statistics for computing the isovector matrix element of the tensor charge for the cB211.072.64 (top), the cC211.060.80 (middle)   and the  cD211.054.96 (bottom) ensemble. The third column gives the number of source positions used. 
      For the cD211.054.96 ensemble, we are increasing statistics for $t_s/a=22,\,24$ and 26. 
      }
      \label{tab:statistics}
  \end{table}
 
\begin{figure}[h!]
    \centering
      \includegraphics[width=0.8\linewidth]{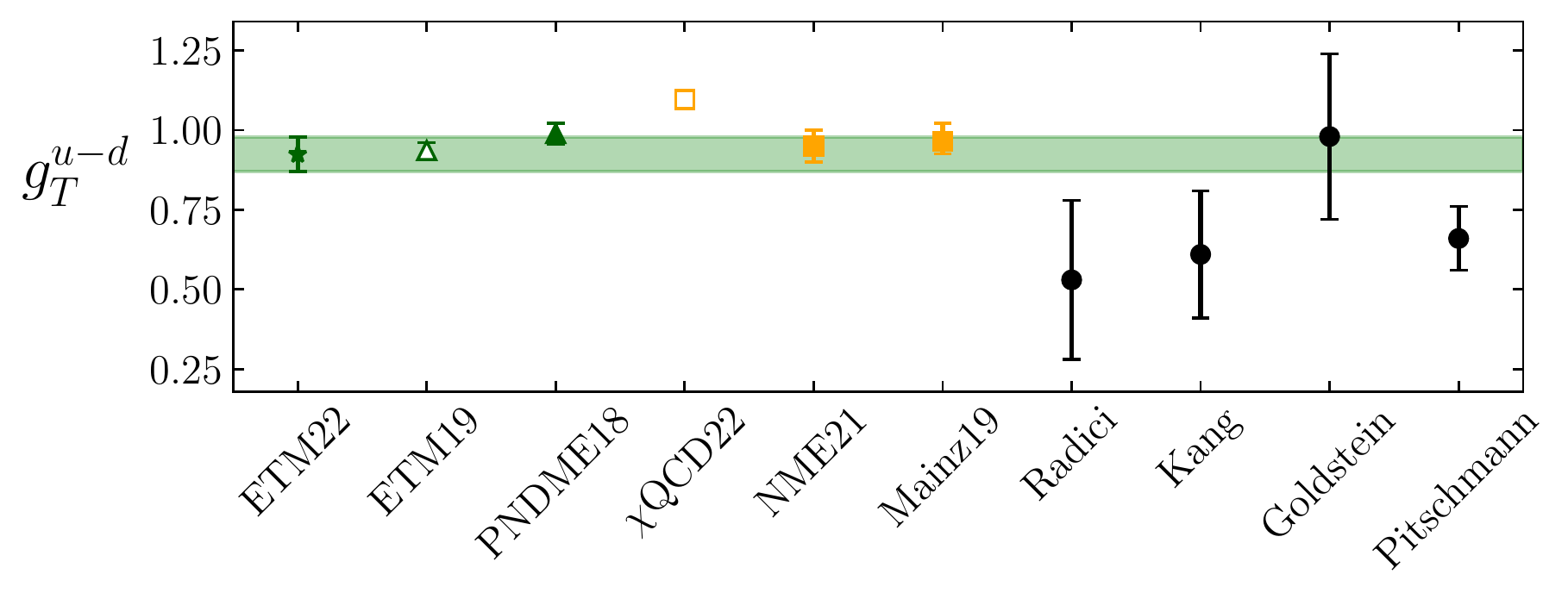}
      \caption{We show lattice QCD results (green symbols for $N_f=2+1+1$ and yellow for $N_f=2+1$ ensembles) for $g_T^{u-d}$ from ETMC~\cite{Alexandrou:2022dtc,Alexandrou:2019brg} using only physical mass point ensembles; PNDME~\cite{Gupta:2018qil}; $\chi$QCD~\cite{Horkel:2020hpi}; NME~\cite{Park:2021ypf}; and CLS-Mainz~\cite{Harris:2019bih}.  Open green and yellow symbols denote lattice QCD results without continuum extrapolation. Results from phenomenology are shown with the black circles~\cite{Radici:2018iag,Kang:2015msa, Goldstein:2014aja,Pitschmann:2014jxa}.\label{fig:gT compare}}\vspace*{-0.5cm}
    \end{figure}

    Using the results for the  three physical mass point  ensembles we extrapolate to the continuum limit avoiding any chiral extrapolations that may introduce unquantified systematic errors.
    We obtain a value of $g_T^{u-d}=0.924(54)$~\cite{Alexandrou:2022dtc}. In Fig.~\ref{fig:gT compare}, we compare our result with those from  other lattice QCD collaborations, all of which have used ensembles simulated with larger than physical pion mass and chirally extrapolated the results to the physical point. As can be seen, the lattice QCD results are in agreement and more accurate than the values extracted from phenomenology, demonstrating that lattice QCD can provide a precise determination.
     
  For the  cB211.072.64 ensemble, we also calculate the tensor charge for the each quark flavor. This requires the computation of disconnected contributions. We use hierachical probing to improve the signal and 600 thousand two-point functions. For the light quark loops we use, in addition,  deflation of the 200 lowest modes. The resulting values are given in Table~\ref{tab:gT q}. 
  \begin{table}
  \centering
\begin{tabular}{l|ccccc}
	\hline
			         & ${u\texttt{-}d}$ & ${u}$  & ${d}$  &     $s$      &     $c$      \\ \hline
			$g_T$    &    0.936(25)     &        0.729(22)       &       -0.207(75)      & -0.00268(58) & -0.00024(16) \\
		\hline
		\end{tabular}
    \caption{We give our values for $g_T$ extracted using the  cB211.072.64 ensemble: first column for $g^{u-d}_T$ and the 2$^{\rm nd}$ to 5$^{\rm th}$ for $g^u_T$, $g_T^d$, $g_T^s$, $g_T^c$, respectively~\cite{Alexandrou:2019brg}. The values   are given in the $\overline{\rm MS}$ scheme at 2~GeV.}\vspace*{-0.3cm}
    \label{tab:gT q}
      \end{table}
     
\begin{figure}[h!]
     \includegraphics[width=\linewidth]{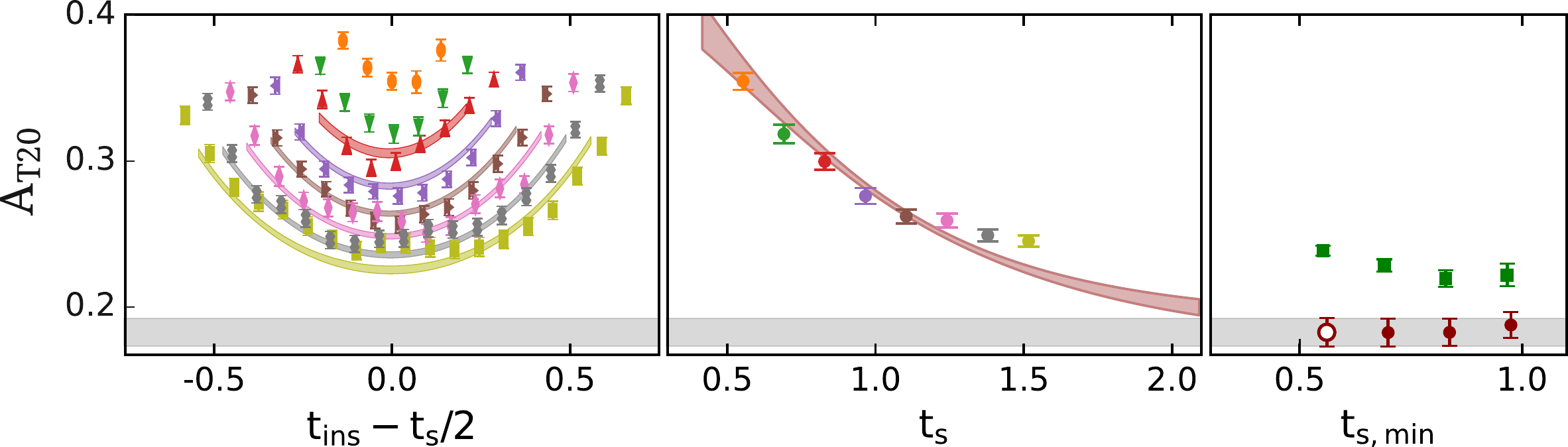}
     \caption{Excited states analysis of the ratio to extract $A_{T20}$ for  the cC211.060.80 ensemble.  The notation is the same as that of Fig.~\ref{fig:gT excited}}
       \label{fig:A20T excited}
   \end{figure}

\begin{figure}[h!]
     \begin{minipage}{0.49\linewidth}
     \includegraphics[width=\linewidth]{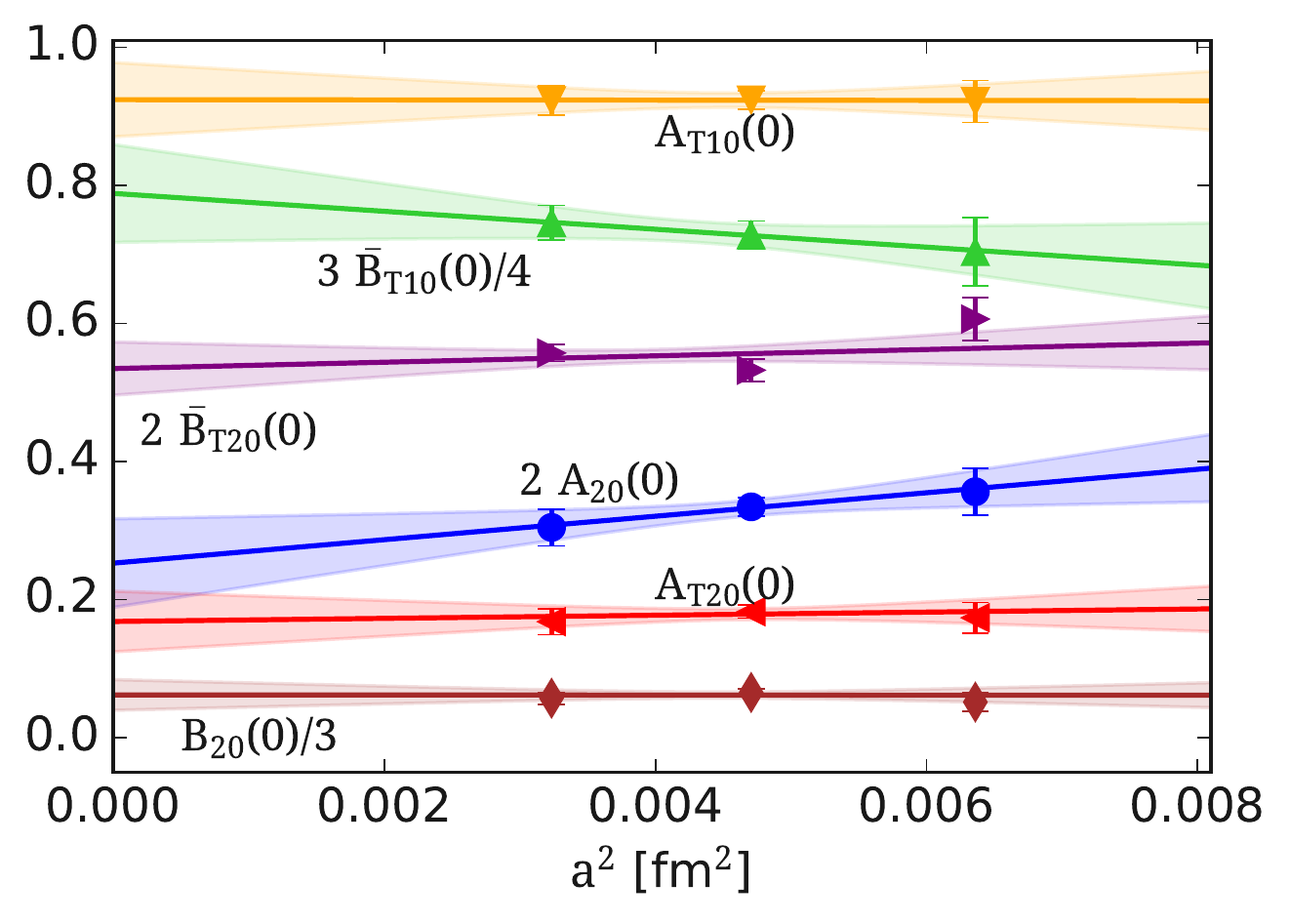}  
     \end{minipage}\hfill
      \begin{minipage}{0.49\linewidth}
     \includegraphics[width=\linewidth]{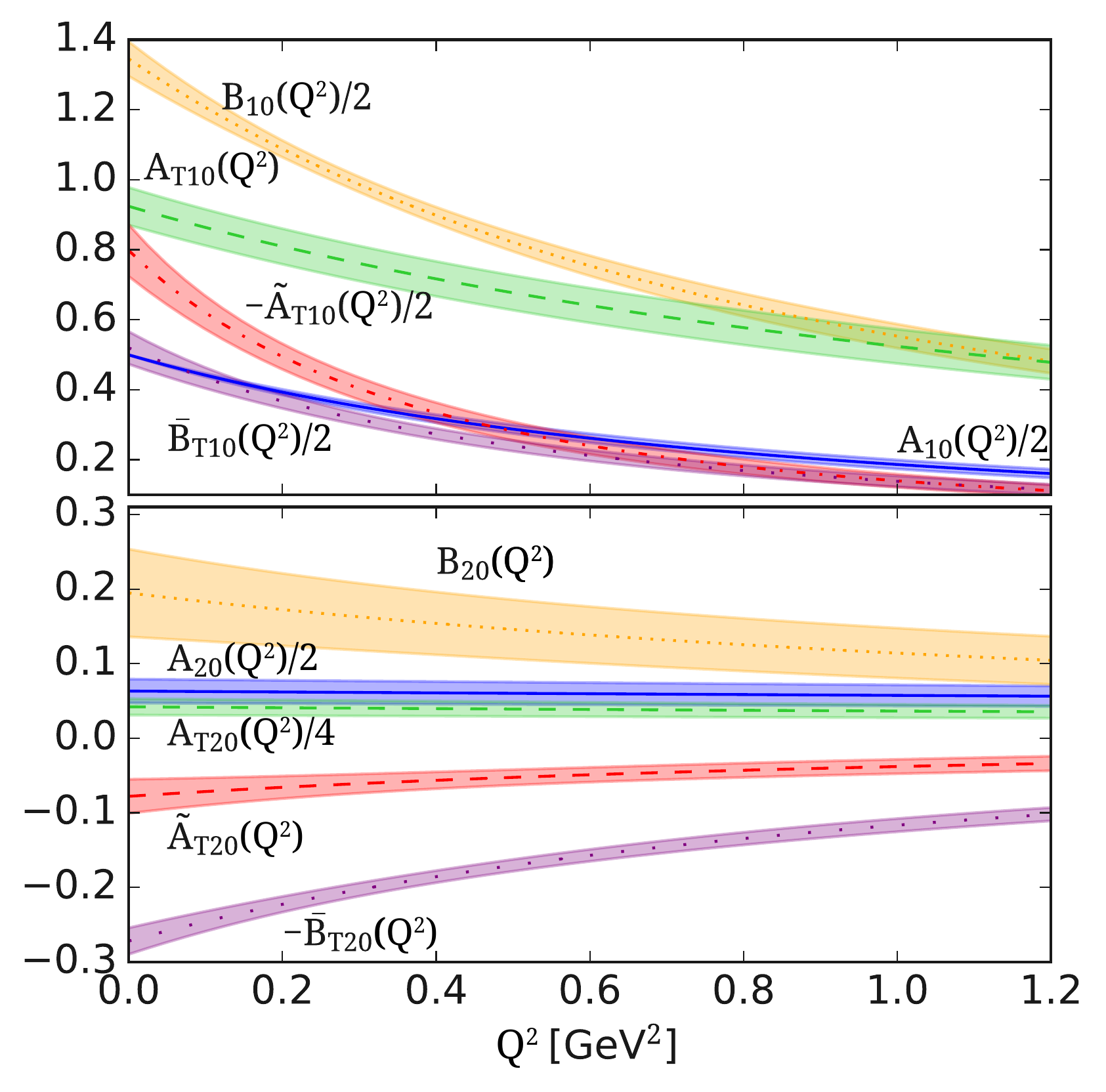}  
      \end{minipage}
      \caption{Left: We show the continuum extrapolation of the isovector GFFs for $Q^2=0$. Right: We show the fits of the $Q^2$-dependence  of the GFFs after extrapolated to the continuum limit. $A_{T10}(0)=g^{u-d}_T$, $A_{20}(0)$ is the isovector momentum fraction and $A_{T20}=\langle x\rangle_{\delta u-\delta d}$.}
      \label{fig:continuum}\vspace*{-0.5cm}
   \end{figure}

  The second Mellin GFFs are extracted by performing a similar analysis. As an example, we show in Fig.~\ref{fig:A20T excited} the extraction of the isovector tensor moment $\langle x \rangle_{\delta u-\delta d}=A^{u-d}_{T20}(0)$ for the cC211.060.80 ensemble.
   Having results for the three ensembles with different lattice spacings, we can perform a continuum extrapolation linearly in $a^2$.  For the GFFs the continuum extrapolation is done at a fixed value of $Q^2$ interpolating the $Q^2$ of two of the ensembles to match the third one. The continuum extrapolations are shown in Fig.~\ref{fig:continuum}.
     \begin{figure}[h!]
     \centering
     \includegraphics[width=0.8\linewidth]{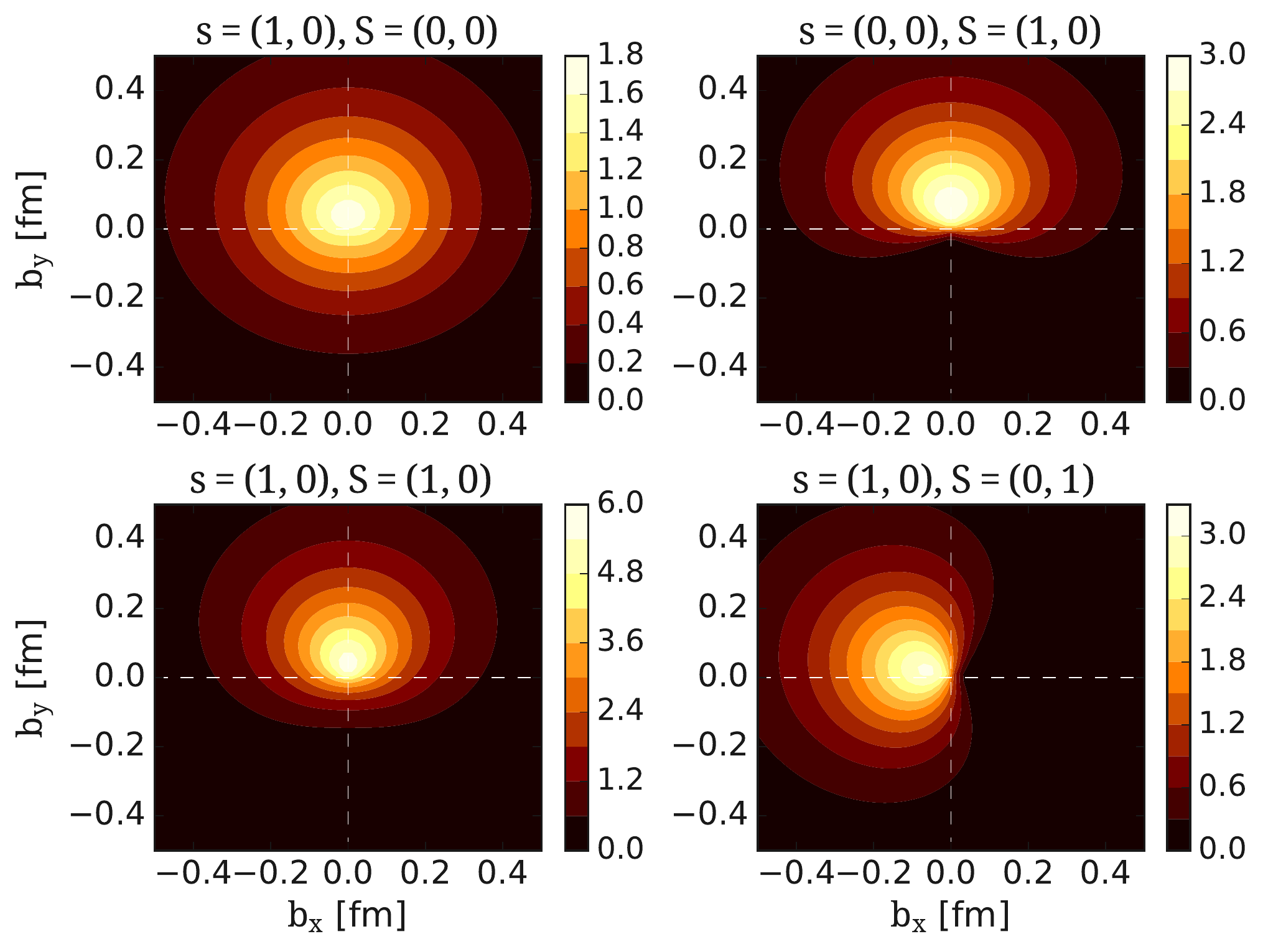}
     \caption{Contours of  $\int_{-1}^1 \rho(x,{\bf b}_\perp ,{\bf s}_\perp ,{\bf S}_\perp)= \langle 1 \rangle_\rho$ [fm$^{-2}$] as a function of $b_x$ and $b_y$. Top-left: transversely polarized quarks in an unpolarized nucleon; top-right: unpolarized quarks in a transversely polarized nucleon; bottom-left: transversely polarized quarks in a transversely polarized nucleon; and bottom-right: transversely polarized quarks in
a perpendicularly polarized nucleon. }\vspace*{-0.3cm}
     \label{fig:x0}
   \end{figure}
   \begin{figure}[h!]
    \centering
     \includegraphics[width=0.8\linewidth]{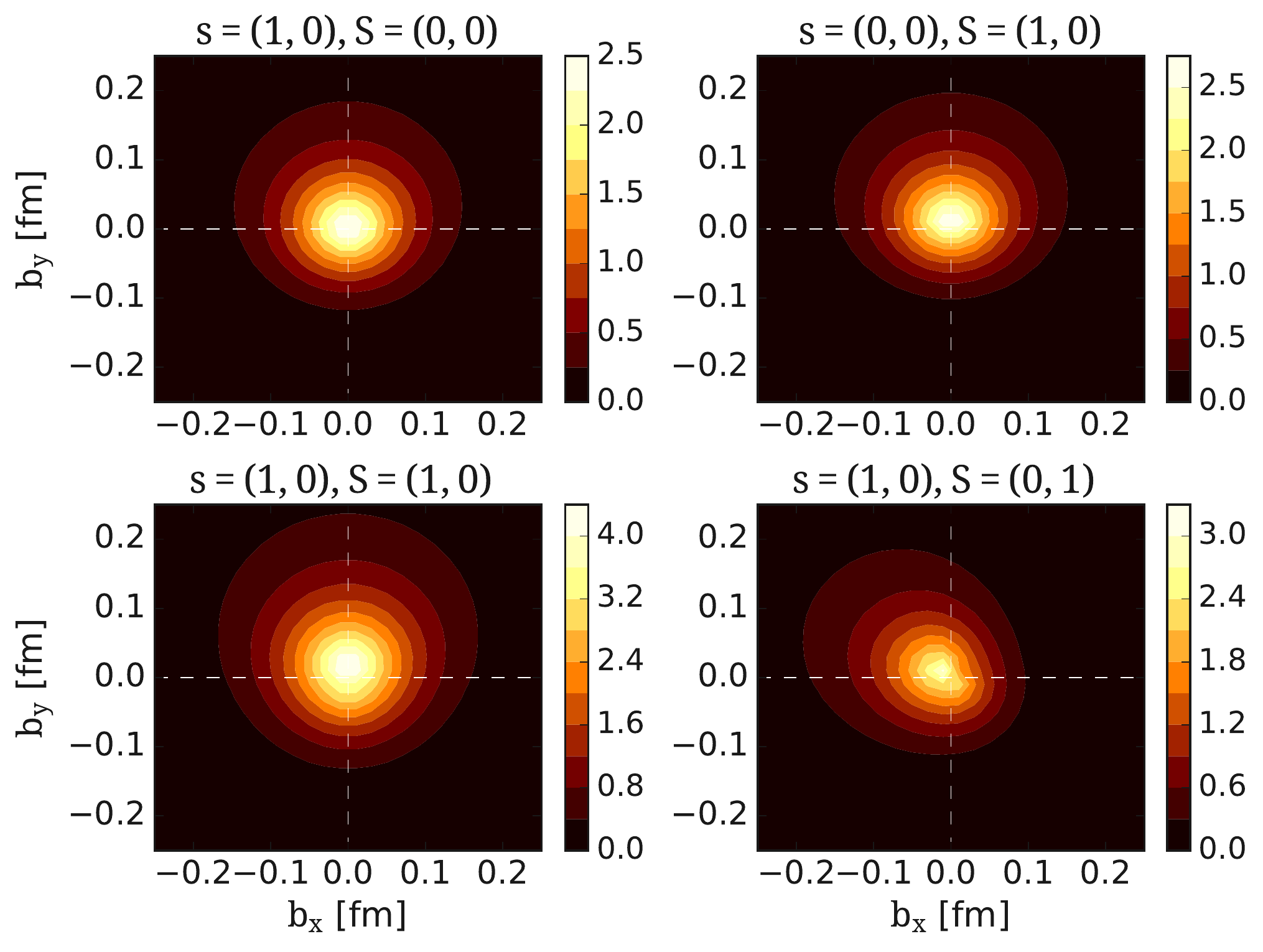}
     \caption{We show  $\int_{-1}^1 \, dx x\rho(x,{\bf b}_\perp ,{\bf s}_\perp ,{\bf S}_\perp)=\langle x \rangle_\rho$ [fm$^{-2}$]. The notation is the same as in Fig.~\ref{fig:x0}.}\vspace*{-0.3cm}
     \label{fig:x1}
   \end{figure}
   We fit the  $Q^2$-dependence of the isovector GFFs after taking the continuum limit  using the form $F(Q^2)=\frac{F(O)}{1+Q^2/m^2)^p}$, where $F(0)$, $m$ and $p$ are fit parameters. We  Fourier transform into impact parameter with the function~\cite{Diehl:2005jf}
   \begin{equation}
     F(b^2_\perp)=\frac{m^2F(0)}{2^p\pi\Gamma(p)}(m b_\perp)^{p-1} K_{p-1}(mb_\perp),
   \end{equation}
   where $\Gamma(x)$ is the Euler gamma function, $K_{n}(x)=K_{-n}(x)$ the modified Bessel functions and $b_\perp = \sqrt{b_\perp^2}$. This then allows us to compute the first and second moments of the transverse density distributions, shown in Figs.~\ref{fig:x0} and \ref{fig:x1}. 
 
   We observe  a sizeable deformation for the first moment. We consider four cases: i) Transversely polarized quarks in an unpolarized nucleon, where we observe a mild distortion because in the isovector combination the $\bar{B}_{T10}(b^2_\perp)$ term  has a milder $Q^2$-dependence; ii) Unpolarized quarks in a transversely polarized nucleon, where we observe a large distortion towards the positive $b_y$ direction. This can be traced back to the GFF $B_{10}$, contributing to the term for $E'$ that, as seen in Fig.~\ref{fig:continuum}, is rapidly decreasing  resulting in a large distortion. The origin of this behavior is related to the Sivers effect~\cite{Bury:2021sue}, a connection that was made in Ref.~\cite{Burkardt:2003uw};  iii) Transversely polarized quarks in
a transversely polarized nucleon, where all the terms in Eq.~(\ref{rho}) contribute  leading to a significant deformation of the   density; and iv) Transversely polarized quarks in
a perpendicularly polarized nucleon, where the third term in Eq.~(\ref{rho}) drops out and the fourth one creates a significant impact, leading to a large distortion  in the $b_x$ direction. For the second moment $\langle x \rangle_\rho$ the distortion in all cases is reduced.

   \section{Conclusions}
   We show that the  moments of PDFs can be extracted precisely and directly at the physical pion mass. Taking the continuum limit at fixed physical pion mass removes any uncontrolled errors due to the chiral extrapolation. The value of the isovector tensor charge of the nucleon provides valuable input to phenomenological analysis, as demonstrated e.g. by the the recent analysis by the JAM collaboration~\cite{Gamberg:2022kdb}, which with lattice input can improve their determination of transverse spin asymmetries.
   The calculation  of sea quark contributions is feasible using physical pion mass ensembles providing, as demonstrated in this work,  the tensor charge for each quark flavor. Although the disconnected contributions are small, they are non-zero for the light, strange and charm quark flavors.
   Using the computation of the unpolarized and transversity GFFs in the continuum limit, one can determine the moments of the transverse quark spin density. We find
large deformation  for the first moment showing a sizeable Sivers effect. We determine the anomalous magnetic moment $\kappa_T\equiv \bar{B}_{T10}(0)=1.051(94)$, a fundamental quantity that   describe the deformation of the transverse polarized quark
distribution in an unpolarized nucleon. First lattice results on $\kappa_T$ were presented by the QCDSF/UKQCD collaboration~\cite{QCDSF:2006tkx}. Since  $\kappa_T \sim - h_1^{\perp}$~\cite{Burkardt:2005hp}, our value confirms  that the Boer-Mulders function, $h_1^{\perp}$, is negative and sizeable and corroborates  the result found in the lattice QCD study of Ref.~\cite{Musch:2011er} for the transverse momentum dependent PDFs.

\begin{acknowledgments}
We would like to thank all members of the Extended Twisted Mass Collaboration  for a very constructive and enjoyable collaboration. M.C. acknowledges financial support by the U.S. Department of Energy Early Career Award under Grant No.\ DE-SC0020405. 
K.H. is supported by the Cyprus Research and Innovation foundation under contract number POST-DOC/0718/0100 and CULTURE-AWARD-YR/0220/0012. S.B., J.F., K.H. and G. S. are partly supported by the EuroCC project (GA No. 951732) 
and by the H2020 project PRACE 6-IP (GA No. 823767).
The project acknowledges support from the European Joint Doctorate projects HPC-LEAP and STIMULATE  funded by the European Union’s Horizon 2020 research and innovation programme under grant agreement No 642069 and 765048, respectively. 
P.D. acknowledges support from the European Union's Horizon 2020 research and innovation programme under the Marie Sklodowska-Curie grant agreement No. 813942 (EuroPLEx) and  from INFN under the research project INFN-QCDLAT.
The authors gratefully acknowledge computing time allocated  on Piz Daint by CSCS 
via the projects with ids s702, s954 and pr79 and on JUWELS at Jülich Supercomputing Centre (JSC) through the projects FSSH, PR74YO and CECY00, CHCH02. Part of the results have been produced within the EA program of JUWELS Booster also with the help of the JUWELS Booster Project Team (JSC, Atos, ParTec, NVIDIA).
We acknowledge the CINECA award under the ISCRA initiative, for the availability of high performance computing resources and support and PRACE for awarding us access to Marconi100 at CINECA (Italy), Piz-Daint at CSCS (Switzerland) and Hawk at HLRS (Germany).
We acknowledge the Gauss Centre for Supercomputing e.V. for project pr74yo  providing computing time on SuperMUC at LRZ and Juwels Booster at JSC. The authors acknowledge the Texas Advanced Computing Center (TACC) at The University of Texas at Austin for providing HPC resources (Project ID PHY21001).
\end{acknowledgments}


\begin{thebibliography}{99}
 \small
\bibitem{Accardi:2012qut}
A.~Accardi, J.~L.~Albacete, M.~Anselmino, N.~Armesto, E.~C.~Aschenauer, A.~Bacchetta, D.~Boer, W.~K.~Brooks, T.~Burton and N.~B.~Chang, \textit{et al.}
Eur. Phys. J. A \textbf{52}, no.9, 268 (2016)
arXiv:1212.1701.\vspace*{-0.2cm} 
  %
\bibitem{Martinelli:1988xs}
G.~Martinelli and C.~T.~Sachrajda,
Phys. Lett. B \textbf{217}, 319-324 (1989).\vspace*{-0.2cm} 
\bibitem{Diehl:2005jf}
M.~Diehl and P.~Hagler,
Eur. Phys. J. C \textbf{44}, 87-101 (2005)
arXiv:hep-ph/0504175.\vspace*{-0.2cm} 
\bibitem{Bar:2016uoj}
O.~B\"ar,
Phys. Rev. D \textbf{94}, no.5, 054505 (2016)
arXiv:1606.09385.\vspace*{-0.2cm} 
\bibitem{Alexandrou:2022dtc}
C.~Alexandrou, S.~Bacchio, M.~Constantinou, P.~Dimopoulos, J.~Finkenrath, R.~Frezzotti, K.~Hadjiyiannakou, K.~Jansen, B.~Kostrzewa and G.~Koutsou, \textit{et al.}
arXiv:2202.09871.\vspace*{-0.2cm} 
\bibitem{Alexandrou:2019brg}
C.~Alexandrou, S.~Bacchio, M.~Constantinou, J.~Finkenrath, K.~Hadjiyiannakou, K.~Jansen, G.~Koutsou and A.~Vaquero Aviles-Casco,
Phys. Rev. D \textbf{102}, no.5, 054517 (2020)
arXiv:1909.00485.\vspace*{-0.2cm} 
\bibitem{Gupta:2018qil}
R.~Gupta, Y.~C.~Jang, B.~Yoon, H.~W.~Lin, V.~Cirigliano and T.~Bhattacharya,
Phys. Rev. D \textbf{98}, 034503 (2018)
arXiv:1806.09006.\vspace*{-0.2cm} 
%
\bibitem{Horkel:2020hpi}
D.~Horkel \textit{et al.} [\ensuremath{\chi}QCD],
Phys. Rev. D \textbf{101}, no.9, 094501 (2020)
arXiv:2002.06699.\vspace*{-0.2cm} 
\bibitem{Park:2021ypf}
S.~Park \textit{et al.} [Nucleon Matrix Elements (NME)],
Phys. Rev. D \textbf{105}, no.5, 054505 (2022)
arXiv:2103.05599.\vspace*{-0.2cm} 
\bibitem{Harris:2019bih}
T.~Harris, G.~von Hippel, P.~Junnarkar, H.~B.~Meyer, K.~Ottnad, J.~Wilhelm, H.~Wittig and L.~Wrang,
Phys. Rev. D \textbf{100}, no.3, 034513 (2019),
arXiv:1905.01291.\vspace*{-0.2cm} 
\bibitem{Radici:2018iag}
M.~Radici and A.~Bacchetta,
Phys. Rev. Lett. \textbf{120}, no.19, 192001 (2018)
arXiv:1802.05212.\vspace*{-0.2cm} 
\bibitem{Kang:2015msa}
Z.~B.~Kang, A.~Prokudin, P.~Sun and F.~Yuan,
Phys. Rev. D \textbf{93}, no.1, 014009 (2016)
arXiv:1505.05589.\vspace*{-0.2cm} 
\bibitem{Goldstein:2014aja}
G.~R.~Goldstein, J.~O.~Gonzalez Hernandez and S.~Liuti,
arXiv:1401.0438.\vspace*{-0.2cm} 
\cite{Pitschmann:2014jxa}
\bibitem{Pitschmann:2014jxa}
M.~Pitschmann, C.~Y.~Seng, C.~D.~Roberts and S.~M.~Schmidt,
Phys. Rev. D \textbf{91}, 074004 (2015)
arXiv:1411.2052.\vspace*{-0.2cm} 
\bibitem{Burkardt:2003uw}
M.~Burkardt,
Nucl. Phys. A \textbf{735}, 185-199 (2004)
arXiv:hep-ph/0302144.\vspace*{-0.2cm} 
\bibitem{Bury:2021sue}
M.~Bury, A.~Prokudin and A.~Vladimirov,
JHEP \textbf{05}, 151 (2021)
arXiv:2103.03270.\vspace*{-0.2cm} 
\bibitem{Gamberg:2022kdb}
L.~Gamberg \textit{et al.} [Jefferson Lab Angular Momentum (JAM) and Jefferson Lab Angular Momentum],
Phys. Rev. D \textbf{106}, no.3, 034014 (2022)
arXiv:2205.00999.\vspace*{-0.2cm} 
\bibitem{QCDSF:2006tkx}
M.~G\"ockeler \textit{et al.} [QCDSF and UKQCD],
Phys. Rev. Lett. \textbf{98}, 222001 (2007)
arXiv:hep-lat/0612032.\vspace*{-0.2cm} 
\bibitem{Burkardt:2005hp}
M.~Burkardt,
Phys. Rev. D \textbf{72}, 094020 (2005)
arXiv:hep-ph/0505189.\vspace*{-0.2cm} 
\bibitem{Musch:2011er}
B.~U.~Musch, P.~Hagler, M.~Engelhardt, J.~W.~Negele and A.~Schafer,
Phys. Rev. D \textbf{85}, 094510 (2012)
arXiv:1111.4249.\vspace*{-0.2cm} 

\end{thebibliography}
\end{document}